\documentclass[graybox,vecphys]{svmult}

\usepackage{mathptmx}       
\usepackage{helvet}         
\usepackage{courier}        
\usepackage{type1cm}        
\usepackage{makeidx}         
\usepackage{graphicx}        
\usepackage{multicol}        
\usepackage[bottom]{footmisc}

\makeindex             

\begin{document}

\title*{Memory effect from spinning unbound binaries}
\author{Lorenzo De Vittori, Achamveedu Gopakumar, Anuradha Gupta, Philippe Jetzer}
\institute{Lorenzo De Vittori, Philippe Jetzer \at Physik-Institut, Universit\"at Z\"urich, Winterthurerstrasse 190, 8057 Z\"urich, Switzerland\\\email{lorenzo@physik.uzh.ch}
\and Achamveedu Gopakumar, Anuradha Gupta \at Department of Astrophysics, Tata Institute of Fundamental Research, Mumbai 400005, India}

\maketitle

\abstract{We present a recently developed prescription to obtain ready--to--use gravitational wave (GW) polarization states for spinning compact binaries on hyperbolic orbits.
We include leading order spin-orbit interactions, invoking 1.5PN-accurate quasi-Keplerian parametrization for the radial part of the orbital dynamics.
We also include radiation reaction effects on $h_+$ and $h_{\times}$ during the interaction.
In the GW signals from spinning binaries there is evidence of the memory effect in both polarizations, in contrast to the non-spinning case, where only the cross polarizations exhibits non-vanishing amplitudes at infinite time.
We also compute 1PN-accurate GW polarization states for non-spinning compact binaries in unbound orbits in a fully parametric way, and compare them with existing waveforms.}

\section{Introduction}
\label{sec:introduction}
Among the plausible sources of GWs for ground and space based detectors, compact binaries moving on hyperbolic orbits could represent an interesting fraction of the observable events.
In the last decades there has been much progress in the description of signals emitted from unbound systems.
Some of the most important works providing crucial inputs required to construct polarization states are the computation of the quadrupolar order radiation field from non-spinning compact objects in Newtonian hyperbolic orbits by Turner \cite{T77}, and its extension to 1PN order by Junker \& Sch\"afer \cite{JS92}, who made use of the quasi-Keplerian parametrization approach by Damour \& Deruelle \cite{DD82} describing 1PN-accurate motion on bound and unbound orbits.
Explicit GW polarization states for non-spinning compact binaries on hyperbolic orbits were computed for the first time by Maj\'ar et al. \cite{MV08,MFV10}.
Many other works provided interesting investigations on different aspects of non-spinning binaries on hyperbolic orbits, such as the energy released during the interactions and the power spectrum of the GW emission \cite{WW76, DKJ12}, or the event rate estimate of such flybys \cite{C08}.

Here we summarize the work done recently by De Vittori et al. in Ref.~\cite{DGGJ14}, where the GW polarization states for spinning compact binaries on hyperbolic orbits were computed, up to 1.5PN order including leading order spin-orbit coupling.
We show the presence of the so-called memory effect, which consists in the non-zero amplitude of the waveform at $t=+\infty$, for both polarization states.

Taking a look at the non-spinning limit, we compute the 1PN-accurate waveforms in a fully parametric manner and show that only the cross polarization exhibits memory effects.
Finally, we compare our signals in this limit case with those in Refs. \cite{JS92, F11} and those obtained through a true anomaly parametrization in Ref.~\cite{MFV10}.

\section{Waveforms for spinning compact binaries on hyperbolic orbits}
\label{sec:spinning}

It is customary to express GWs polarization states at Newtonian order through the following expression in terms of the velocity vector $\vec{v}$, the unit relative separation vector $\vec{n}$ and the plane--of--the--sky unit vectors $\vec{p}$ and $\vec{q}$:
\begin{eqnarray}
  \label{eq_h+}
  h_{+}& = &\frac{2 G \mu}{c^4 R}\bigg\{ (\vec{p} \cdot \vec{v} )^2 - (\vec{q} \cdot \vec{v} )^2 - \frac{Gm}{r} \left[ ( \vec{p} \cdot \vec{n} )^2 - ( \vec{q} \cdot \vec{n} )^2 \right] \bigg\}~, \\
  \label{eq_hx}
  h_{\times}& = &\frac{4 G \mu}{c^4 R}\bigg\{(\vec{p} \cdot \vec{v}) \,(\vec{q} \cdot \vec{v}) - \frac{Gm}{r}\,(\vec{p} \cdot \vec{n})\, (\vec{q} \cdot \vec{n}) \bigg\}~,
\end{eqnarray}
where $m$ is the total mass of the system, $\mu$ the reduced mass, $R$ the distance to the observer and $r$ the relative separation.
It is possible to express the polarization states for general orbits including leading order spin-order coupling contributions up to 1PN, as given e.g. in Ref.~\cite{GS11}.
Choosing the reference frames in an appropriate way and performing some straightforward computations, one can express the above vectors and dot products in terms of the dynamics,
i.e. as a function of the relative radius $r$, the radial velocity $\dot r$, the pahsing angle $\Phi$ and its derivative $\dot \Phi$, and the precession angles $\alpha$ and $\iota$,
obtaining 
\begin{equation}\label{eq:h_dyn}
  h_+|_Q(t) = h_+(r,\dot{r},\Phi,\dot{\Phi},\alpha,\iota)\quad {\rm and} \quad  h_{\times}|_Q(t) = h_{\times}(r,\dot{r},\Phi,\dot{\Phi},\alpha,\iota)~.
\end{equation}
Explicit expressions can be found in Sec. II of Ref.~\cite{DGGJ14}.

Since we are interested in temporally evolving $h_+|_Q(t)$ and $h_{\times}|_Q(t)$, we need to specify how $r$, $\dot r$, $\Phi$, $\dot \Phi$, $\alpha$ and $\iota$ vary in time along PN-accurate hyperbolic orbits.
In the case of spinning binaries, we have to divide this analysis into two parts, the radial and the angular solution of the dynamics.
The former can be solved in a customary parametric way, while the latter has to be tackled numerically.

Using expressions for non-spinning binaries on eccentric PN-accurate orbits from Ref.~\cite{DD82}, results including spin effects were found by Gopakumar \& Sch\"afer \cite{GS11}.
They solve the radial part of the dynamics through the usual Kepler equations
\begin{eqnarray}
  r& = &a_{\rm r} \;(1-e_{\rm r}\, \cos u)~,\label{eq:r_ecc_kepler}\\
  l& = &n\;(t-t_0) = u - e_{\rm t} \sin u~, \label{eq:rdot_ecc_kepler}
\end{eqnarray}
where $a_{\rm r}$, $e_{\rm r}$, $e_{\rm t}$ and $n$ are the PN extensions of the semi-major axis, the radial eccentricity, the time eccentricity and the mean motion, respectively.
Afterwards, they find expressions for the orbital parameters $a_{\rm r}$, $e_{\rm t}$ and $e_{\rm r}$ for eccentric orbits, as functions of the dimensionless parameter $\xi = Gmn/c^3$, the reduced angular momentum $L$ and energy $E$, the symmetric mass ratio $\eta$, and the spin-orbit interaction parameter $\Sigma$, describing the spin misalignment in terms of $\vec{k}\cdot\vec{s}_1$ and $\vec{k}\cdot\vec{s}_2$.
Finally, they can write down the solution to equation (\ref{eq:r_ecc_kepler}) as a series in $1/c$ up to 1.5PN order.

We obtain the hyperbolic counterpart of the Kepler equations and their solutions invoking the analytic continuation technique from Ref.~\cite{DD82}, i.e. we let $u=\imath v$ and $n=-\imath \bar n$.
The analog of eqs. (\ref{eq:r_ecc_kepler}) and (\ref{eq:rdot_ecc_kepler}) for unbound orbits is
\begin{eqnarray}
  r& = &a_{\rm r}\;(e_{\rm r}\,\cosh v -1)~,\\
  l& = &\bar n\;(t-t_0) = e_{\rm t} \sinh v - v~,
\end{eqnarray}
and computing the hyperbolic version of the orbital parameters, the radial solution can be expressed through $\bar \xi = Gm\bar n/c^3$ as
\begin{eqnarray}
  r& = &\frac{Gm}{c^2}\frac{1}{\bar \xi^{2/3}}\bigg\{e_{\rm t} \cosh{v} -1 - \bar \xi^{2/3}\frac{18-2\eta+(6-7\eta)\,e_{\rm t}\cosh{v}}{6}+\frac{\bar \xi\;\Sigma}{\sqrt{e_{\rm t}^2-1}}\bigg\}~,\quad\label{eq:r}\\
  \dot{r}& =\; &\bar \xi^{1/3}\,\frac{c\,e_{\rm t}\,\sinh{v}}{e_{\rm t}\cosh{v}-1}\;\bigg\{1-\bar \xi^{2/3}\,\frac{6-7\eta}{6}\bigg\}~.\label{eq:rdot}
\end{eqnarray}

For the angular part of the dynamics, we employ again the results in Ref.~\cite{GS11} for eccentric orbits.
Using the previous arguments to find the hyperbolic counterpart, we find the differential equations driving the temporal evolution:
\begin{eqnarray}
  \dot{\vec{s_1}}& = &\frac{c^3}{Gm}\frac{\bar \xi^{5/3}\sqrt{e_{\rm t}^2-1}}{(e_{\rm t}\cosh v-1)^3}\;\delta_1\;\vec{k}\times\vec{s}_1~,\label{eq:dot_phi}\\
  \dot{\vec{s_2}}& = &\frac{c^3}{Gm}\frac{\bar \xi^{5/3}\sqrt{e_{\rm t}^2-1}}{(e_{\rm t}\cosh v-1)^3}\;\delta_2\;\vec{k}\times\vec{s}_2~,\\
  \dot{\vec{k}}& = &\frac{c^3}{Gm}\frac{\bar \xi^2\;(\delta_1\chi_1q\;\vec{s}_1+\delta_2\chi_2/q\;\vec{s}_2)\times\vec{k}}{(e_{\rm t}\cosh v-1)^3}~,\\
  \dot \Phi& =\, &\frac{\bar n\,\sqrt{e_{\rm t}^2-1}}{(e_{\rm t}\cosh v -1)^2}\bigg\{1+\bar \xi^{2/3}\left(\frac{4-\eta}{e_{\rm t}\cosh v -1}+\frac{\eta-1}{e_{\rm t}^2-1}\right)\,-\nonumber\\
	   &&\bar \xi \,\frac{\Sigma}{\sqrt{e_{\rm t}^2-1}} \left(\frac{1}{e_{\rm t}\cosh v -1}+\frac{1}{e_{\rm t}^2-1}\right)\!\bigg\}-\cos\iota\;\dot{\alpha}~,\label{eq:phi_dot}
\end{eqnarray}
where the derivative in the last term of eq. (\ref{eq:phi_dot}) reads $\dot \alpha = (k_x \dot{k}_y - k_y \dot{k}_x)/(k_x^2 + k_y^2)$.
Additionally, it is possible to incorporate numerically the effects of radiation reaction during the hyperbolic passage.
To the above set, we add the following 2.5PN-accurate coupled differential equations for $\bar n$ and $e_{\rm t}$, adapted from \cite{DGI04}:
\begin{eqnarray}
\frac{d\bar n}{dt}&  = &-\frac{c^6}{G^2m^2}\frac{\bar \xi^{11/3}\;8\;\eta}{5\;\beta^7}\,\left(-49\beta^2-32\beta^3+35(e_{\rm t}^2-1)\beta-6\beta^4+9e_{\rm t}^2\beta^2\right)~,\\
  \frac{de_{\rm t}}{dt}&  = &-\frac{c^3}{Gm}\frac{\bar \xi^{8/3}8\eta(e_{\rm t}^2-1)}{15\;\beta^7\;e_{\rm t}}\left(-49\beta^2-17\beta^3+35(e_{\rm t}^2-1)\beta-3\beta^4+9e_{\rm t}^2\beta^2\right)~,\qquad\label{eq:et_dot}
\end{eqnarray}
where for simplicity $\beta = e_{\rm t}\cosh v -1$.
Specifying the initial binary configuration in terms of the masses $m_1$, $m_2$, the Kerr parameters $\chi_1$, $\chi_2$, the initial values for $\bar n$ and $e_{\rm t}$
and the initial spin orientations through the angles ($\theta_1^i$, $\theta_2^i$) and ($\phi_1^i$, $\phi_2^i$),
we implement a numerical method to evolve the set of 12 differential equations (\ref{eq:dot_phi}) to (\ref{eq:et_dot}),
and combine it to the results for the radial part in equations (\ref{eq:r}) and (\ref{eq:rdot}).
Finally, we impose these variations in the expressions for $h_+|_Q(t)$ and $h_{\times}|_Q(t)$, given in (\ref{eq:h_dyn}),
and obtain the desired waveforms for spinning compact binaries on hyperbolic orbits.

In Fig.~1 we display the waveform for both polarization states, resulting from such an implementation.
We compare the signals for two different initial eccentricities $e_{\rm t} = 1.5$ and $e_{\rm t} = 1.3$, and for two different mass ratios $q=1$ and $q=4$.
\begin{figure}
  \label{fig:spinning}
  \includegraphics[width=\textwidth]{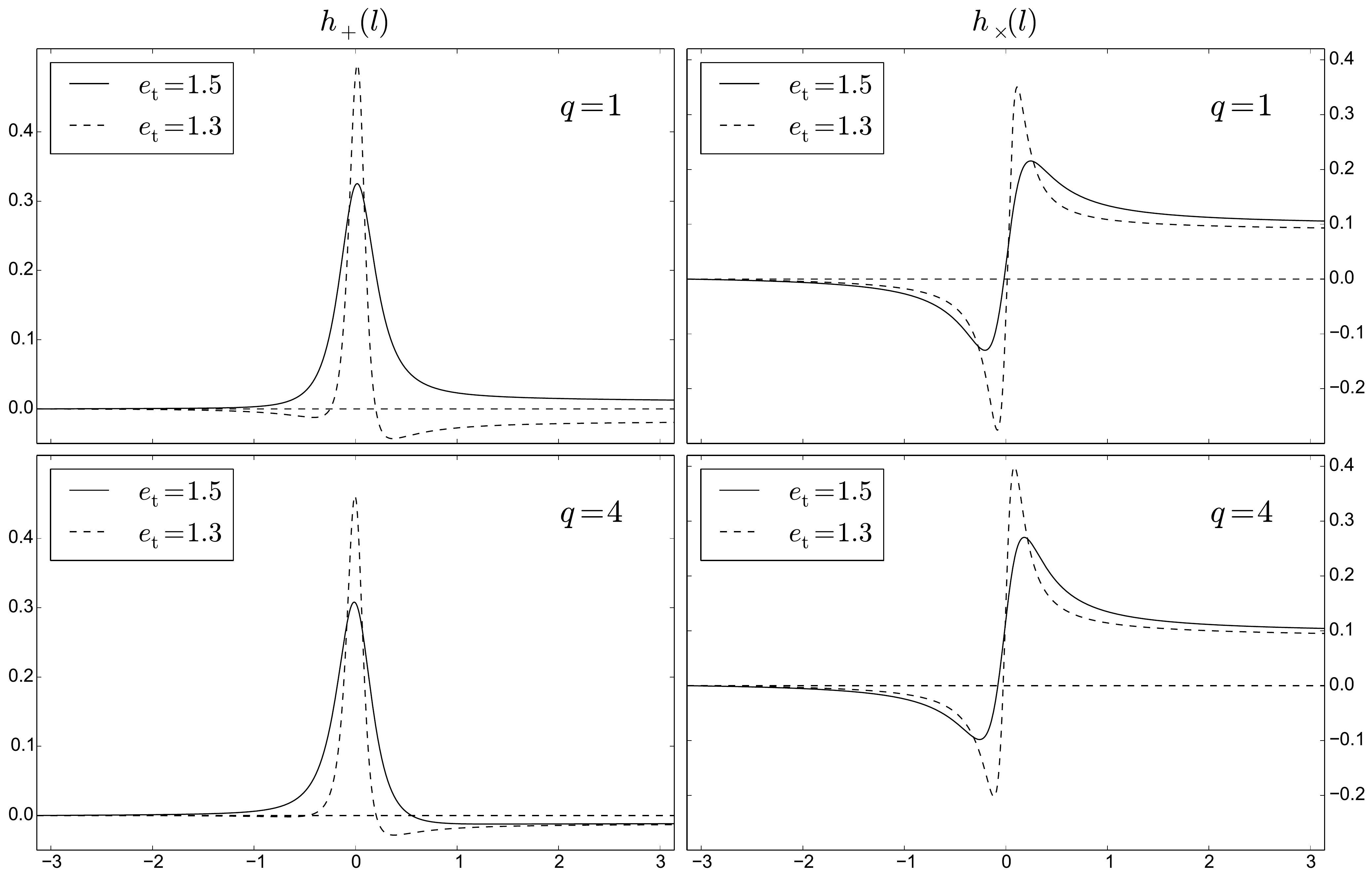}
  \caption{We plot $h_+|_Q$ and $h_{\times}|_Q$ for mass ratios $q=1$ and $q=4$.
  As one can see, both polarizations show a memory effect: the amplitude of the waveform at $t=+\infty$ is different from the initial.
  The hyperbolic passage generates a burst and leaves a trace of the interaction in the metric.
We see the influence of the eccentricity on the waveform by comparing the $e_{\rm t} = 1.5$ and $e_{\rm t} = 1.3$ cases.}
\end{figure}

\section{Waveforms for non-spinning binaries on hyperbolic orbits}
\label{sec:non-spinning}

In the limit case of non-spinning compact binaries it is easier to obtain PN-accurate waveforms since there is no spin-orbit coupling and therefore no precession of the orbital plane.
It is then enough to find the evolution to the radius $r$ and its derivative $\dot r$, as well as the phasing angle $\phi$ and its derivative $\dot \phi$ to fully describe the dynamics of the binary.
This can be done in a parametric manner, including 1PN correction terms, as done in Ref.~\cite{DD82}. Results for $r$, $\dot r$, $\phi$, $\dot \phi$ can be found in Sec. III of \cite{DGGJ14}.

The expressions for the polarization states we use are the same as for the spinning case, whereas the dot products in (\ref{eq_h+}-\ref{eq_hx}) and in their 1PN generalization are much simpler.
As usual, we can write them as a PN series:
\begin{equation}
  h_+ = -\frac{G\mu}{c^4R}\, \left ( h_+^N + \frac{1}{c}h_+^{0.5} + \frac{1}{c^2}h_+^{1} \right )~, \quad h_{\times} = -\frac{G\mu}{c^4R}\, \left ( h_{\times}^N + \frac{1}{c} h_{\times}^{0.5} + \frac{1}{c^2} h_{\times}^{1} \right )~.
\end{equation}
In Sec. III of Ref.~\cite{DGGJ14} we show the full 1PN expressions in terms of $r$, $\dot r$, $\phi$ and $\dot \phi$.

Specifying the initial configuration of the binary and imposing the evolution equations in the above expressions, we obtain waveforms for non-spinning compact binaries on hyperbolic orbits.

In Fig.~2 we plot some results for a binary system with masses $m_1 = 8M_{\odot}$ and $m_2 = 13M_{\odot}$, for $e_{\rm t} = 2$ and $e_{\rm t} = 1.7$, showing the waveform at different PN orders.
\begin{figure}
  \label{fig:non-spinning}
  \includegraphics[width=\textwidth]{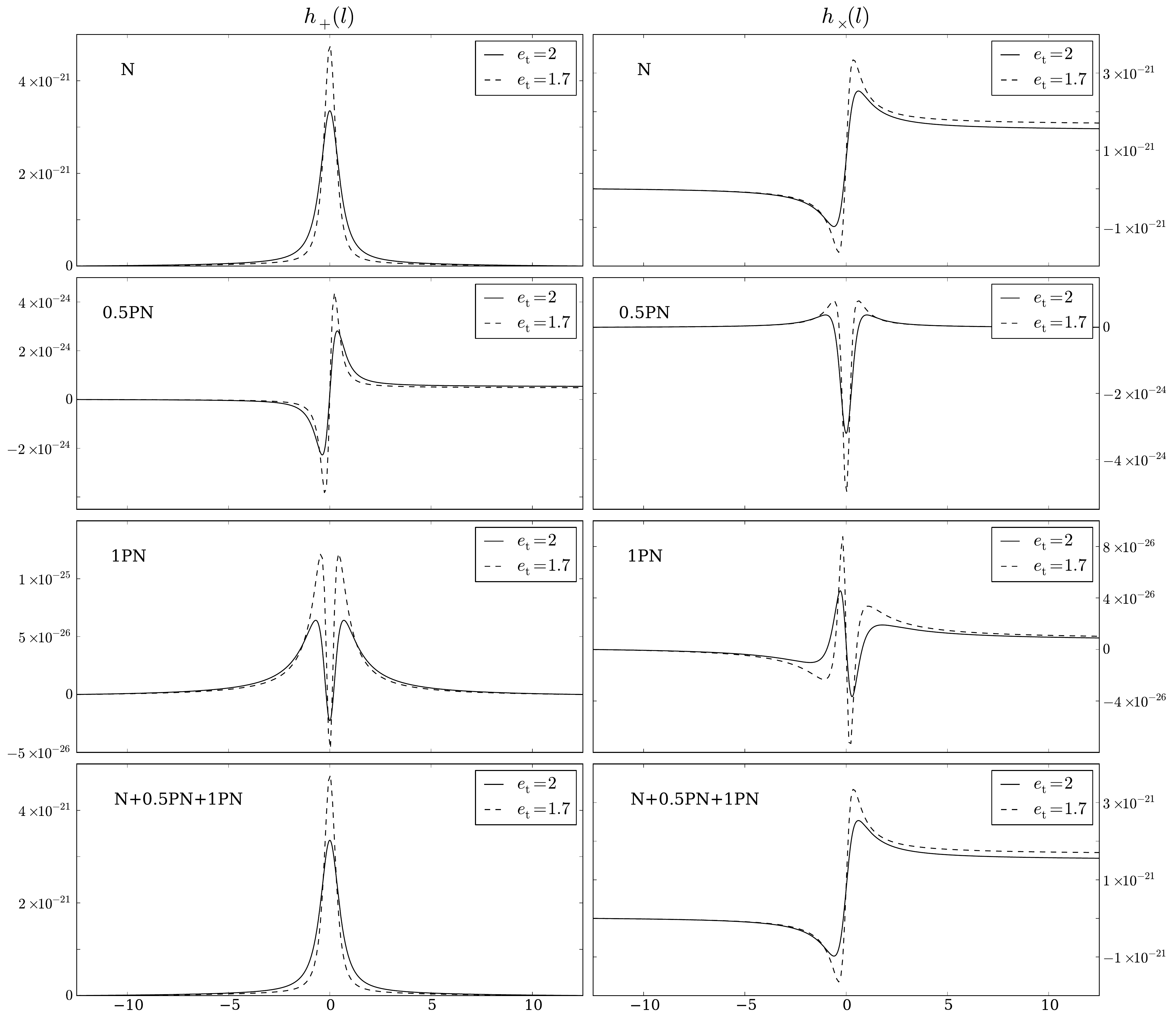}
  \caption{We plot here $h_+|_Q$ and $h_{\times}|_Q$ for non-spinning binaires, at Newtonian, 0.5PN and 1PN order.
  In contrast to the spinning case, here only the cross polarization shows a memory effect.
  Also, one can see that the signal is symmetric with respect to the $t=0$ axis at each PN order.
  This is due to the fact that the orbital plane is not precessing.
  The effect of the eccentricity on the waveform is clearly visible.
  Low eccentricities lead to violent interactions and therefore sharp and high peaks.}
\end{figure}

We can compare our results for non-spinning binaries along unbound orbits with existing waveforms.
We have reproduced temporal evolution for the real and the imaginary parts of appropriate time derivatives of various mass and current multipole moments that are displayed in Fig.~8 in \cite{JS92}, so that we can be sure of the correctness of our approach. Also, results depicted in Fig.~2 of Ref.~\cite{F11} for the (2,2) GW mode coincide with ours.

In Ref.~\cite{MV08}, a generalized true anomaly parametrization approach to obtain 1PN-accurate polarization states is developed.
The prescpription is carried out in Ref.~\cite{MFV10} and some results are shown.
A visual comparison of our $e_{\rm t} = 2$ plots in Fig.~2 with Figs. 6, 7 and 10 of Ref.~\cite{MFV10} reveals considerable differences.
The memory effect exhibited in the cross polarization is clearly different,
and none of the GW signals depicted seems to be symmetric with respect to the $t=0$ axis, something we would expect.
We suspect that the observed differences may be due to the way temporal evolution is implemented in Ref.~\cite{MFV10}.

\section{Conclusions}
\label{sec:conclusion}

We provide a new prescription to obtain ready--to--use PN-accurate GW templates for spinning compact binaries on hyperbolic orbits,
including leading order spin-orbit coupling contributions and radiation reaction effects.
We are able to reproduce previous results for the non-spinning limit case, and find some interesting new feature arising from the spin-orbit coupling.
In particular, we notice that for spinning binaries both the plus and the cross polarization states exhibit a memory effect, i.e. a non-vanishing amplitude at time $t=+\infty$,
whereas for non-spinning systems only the cross polarization shows such a behaviour. In this last case, we can write
\begin{equation}
  \lim_{t\to +\infty} h_+|_Q(t) = \lim_{t\to -\infty} h_+|_Q(t)~,\qquad \lim_{t\to +\infty} h_{\times}|_Q(t) = - \lim_{t\to -\infty}h_{\times}|_Q(t)~,
\end{equation}
while for spinning binaries not only the sign but also the magnitude of $h_+,{\times}$ changes from $t=-\infty$ to $t=+\infty$.
Since the waveform is strongly influenced by spin effects, the structure of such an asymmetric burst-like signal carries many informations about the system which could be useful for parameter estimation.
It will be interesting to incorporate the 2PN order non-spinning contributions to our 1.5PN-accurare orbital dynamics, as well as dominant order spin-spin interactions.

%
%

\begin{thebibliography}{99.}%

\bibitem{T77} M. Turner, Astrophys. J. \textbf{216}, 914 (1977)
\bibitem{JS92} W. Junker and G. Sch\"afer, Mon- Not. R. Astr. Soc. \textbf{254}, 146 (1992)
\bibitem{DD82} T. Damour and N. Deruelle, Ann. Inst. Henri Poincar\'e Phys. Th\'eor. \textbf{43}, 107 (1985)
\bibitem{MV08} J. Maj\'ar and M. Vas\'uth, Phys. Rev. D \textbf{77}, 104005 (2008)
\bibitem{MFV10} J. Maj\'ar, P. Forg\'acs and M. Vas\'uth, Phys. Rev. D \textbf{82}, 064041 (2010)
\bibitem{WW76} R. V. Wagoner and C. M. Will, Astrophys. J. \textbf{210}, 764 (1976)
\bibitem{DKJ12} L. De Vittori, A. Klein and P. Jetzer, Phys. Rev. D \textbf{86}, 044017 (2012)
\bibitem{C08} S. Capozziello, M. de Laurentis, F. de Paolis, G. Ingrosso and A. Nucita, Modern Physics Letters A \textbf{23}, 99 (2008)
\bibitem{DGGJ14} L. De Vittori, A. Gopakumar, A. Gupta and P. Jetzer, submitted to Phys. Rev. D. arXiv pre-print: 1410.6311
\bibitem{F11} M. Favata, Phys. Rev. D \textbf{84}, 124013 (2011)
\bibitem{GS11} A. Gopakumar, G. Sch\"afer, Phys. Rev. D \textbf{84}, 124007 (2011)
\bibitem{DGI04} T. Damour, A. Gopakumar, and B. R. Iyer, Phys. Rev. D \textbf{70}, 064028 (2004)

\end{thebibliography}

\end{document}